\documentclass[a4paper]{jpconf}
\usepackage{graphicx}
\newcommand{\pp}{$p$-$p$}
\def\ttt#1{\texttt{\small #1}}

\begin{document}
\title{Studies of isolated photon production in simulated proton-proton collisions with ALICE-EMCal}

\author{Rapha\"elle Ichou for the ALICE Collaboration}

\address{Subatech, 4 rue Alfred Kastler, BP 20722 44307 NANTES cedex 3, France}

\ead{raphaelle.ichou@cern.ch}

\begin{abstract}
The production of prompt photons at high transverse momentum in proton-proton collisions (\pp) is a useful tool to study perturbative Quantum-Chromo-Dynamics (pQCD). In particular, they yield valuable information about parton distribution functions in the proton. 
The experimental measurement of prompt $\gamma$ is a difficult task due to the large background of decay photons from neutral mesons, mainly $\pi^{0}$.
We present a full simulation and reconstruction study of prompt $\gamma$ identification in \pp\ at $\sqrt{s}= 14$ TeV in the ALICE electromagnetic calorimeter EMCal, giving details on the methods developed to separate them from decay photons with the help of shower-shape and isolation cuts. We present Monte Carlo predictions for signal and background. The method used to extract the final isolated $\gamma$ corrected cross-section is presented and the calculation of various experimental corrections is outlined.
\end{abstract}

\section{Physics motivations}

The study of photon production at large transverse momenta ($p_T \gg \Lambda_{QCD}$ = 0.2 GeV) in hadronic interactions is a valuable testing ground of the perturbative regime of Quantum Chromodynamics (pQCD)~\cite{Aurenche:2006vj}. At the LHC, photons will allow one to confront the data with pQCD predictions at energies never reached before. Since these photons come directly from parton-parton hard scatterings, they allow one to constrain the gluon distribution function in the proton at small parton momentum fraction $x$ = $p_{parton}$/$p_{proton}$. Also, photons produced in \pp\ collisions provide a Ç vacuum È baseline reference for the study of their production rates in nucleus-nucleus collisions. \\
At lowest order, 
three partonic mechanisms produce prompt $\gamma$ in hadronic collisions:
(i) quark-gluon Compton scattering $q g \rightarrow \gamma q$, (ii) quark-antiquark annihilation $q \bar q \rightarrow \gamma g$, and (iii) the collinear fragmentation of a final-state parton into a photon, e.g. $qq\rightarrow qq \rightarrow \gamma X$. The photons produced in the two first point-like processes are called \textit{direct}, the latter \textit{fragmentation} photons. The Compton channel is particularly interesting as it provides information on the proton gluon distribution~\cite{Ichou:2010wc}, which is otherwise only indirectly constrained via the ``scaling violations" of 
the proton structure function 
in deep-inelastic scattering (DIS) $e$-$p$ collisions~\cite{Klein:2008di}.


The measurement of high-$p_T$ photon production is complicated by a large $\gamma$ background from hadrons, specially from $\pi^0$ mesons, which decay into two photons.
At high $p_T$ there are between 10 and 100 (at 100~GeV/c and 10~GeV/c respectively) times more $\pi^0$ than prompt $\gamma$. Furthermore, above 30~GeV/c, the two $\pi^0$ decay photons merge into a single cluster in EMCal and thus cannot be identified as a $\pi^0$ via $\gamma$-$\gamma$ invariant mass analysis.
In order to measure the prompt $\gamma$ signal out of the overwhelming background of $\pi^0$, which are produced in the fragmentation of jets, one requires the photon candidate to be isolated from any hadronic activity within a given distance around its direction. The corresponding measurements are then dubbed \textit{isolated} photons. A standard isolation requirement is that within a cone around the $\gamma$ direction defined in pseudo-rapidity $\eta$ and azimuthal angle $\phi$ by $R = \sqrt{(\eta - \eta_{\gamma})^2 + (\phi - \phi_{\gamma})^2}$, the accompanying hadronic transverse energy is less than a fixed fraction $\varepsilon$ (e.g. often 10\%) of the photon's $p_T$. $R$ is usually taken between 0.4 and 0.7. 
Isolation enables one to reject a large fraction of the $\pi^0$ decay photons, without supressing too many photons of interest.
The bigger $R$ and the smaller $\varepsilon$, the more selective is the isolation. But the choice of optimal isolation criteria also needs to consider detector acceptance and Underlying Event (UE) limitations. The values choosen are $R =0.2$ (for the first year of running at $\sqrt{s}$~= 7~TeV, with only 4 Super-Modules of EMCal installed), $R=0.4$ and $\varepsilon$ = 0.1 (which e.g. removes UE-hadrons with $p_T<$2 GeV/c in events with a 20-GeV $\gamma$).

\section{Monte Carlo predictions}

The \ttt{JETPHOX}~\cite{jetphox} program (for JET-PHOton/hadron X-sections) allows one to calculate perturbatively the $\gamma$ cross-sections at Next to Leading Order accuracy. It includes both the fragmentation and the direct components and isolation cuts can be applied at the parton level.
\begin{figure}[htbp!]
\begin{center}
    \includegraphics[height=7.8cm, width=7cm, clip=true]{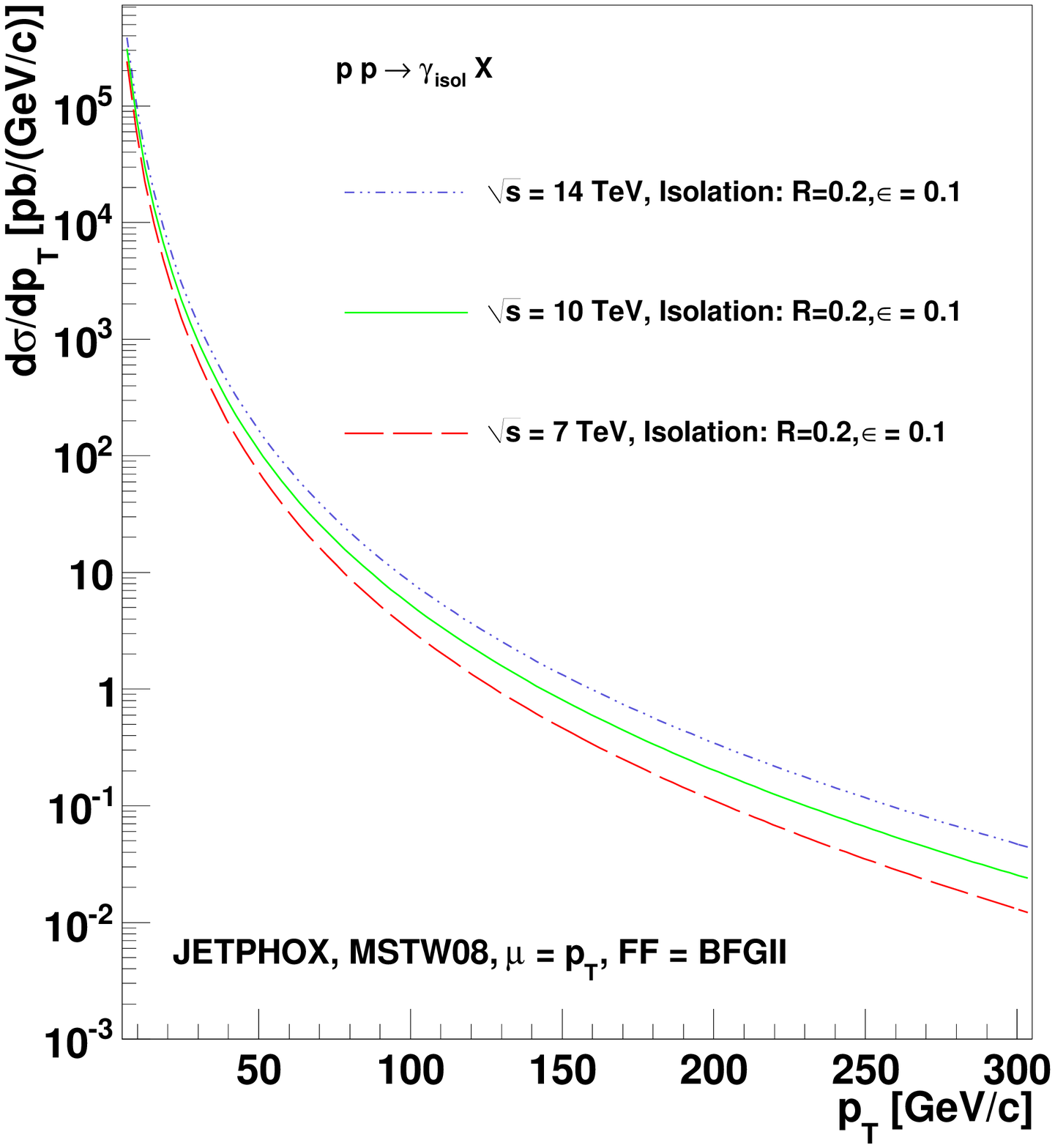}
    \includegraphics[height=7.8cm, width=7cm, clip=true]{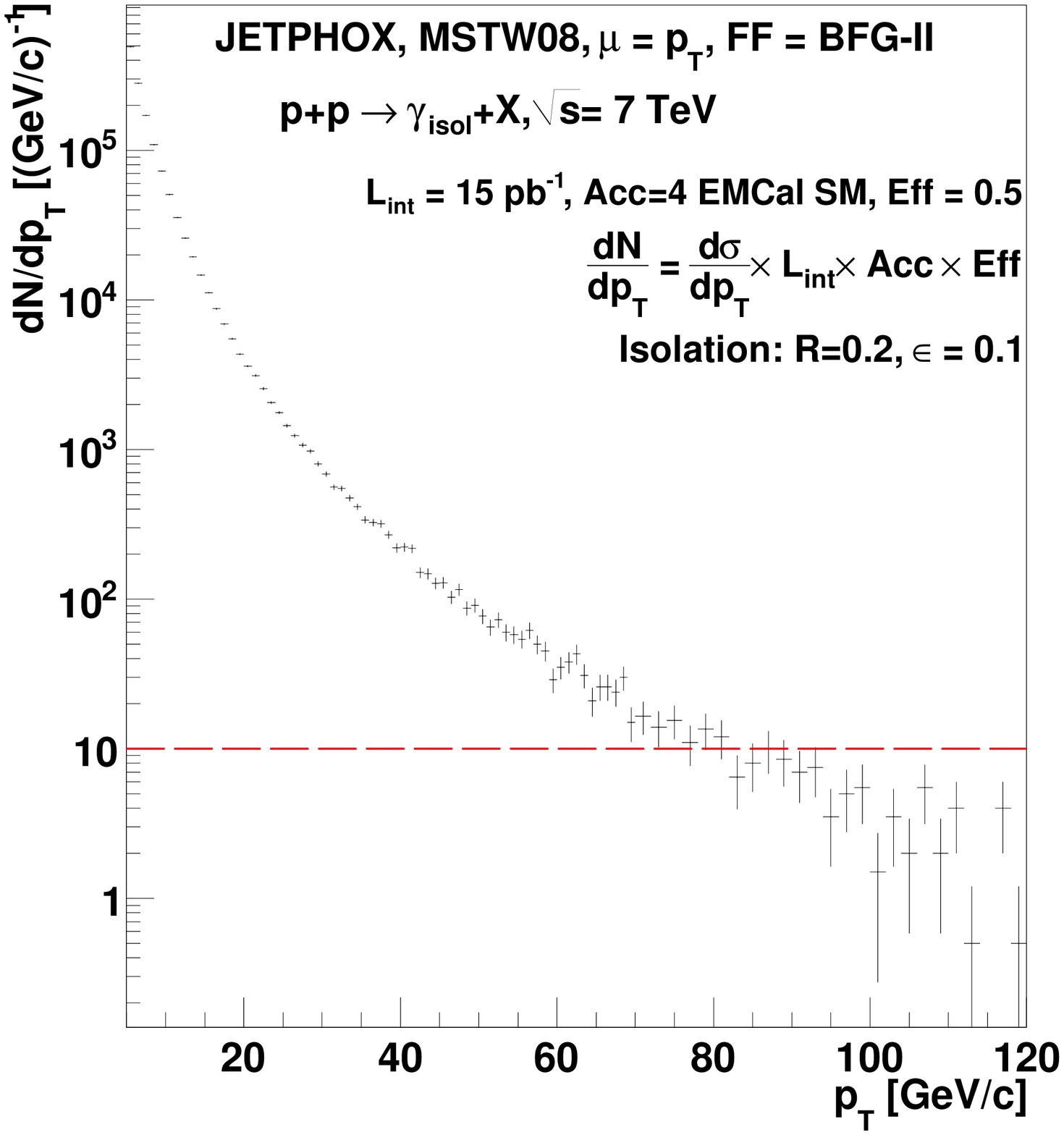}
          \caption{Differential cross-section as a function of $p_T$ for isolated photons ($R$~=0.2, $\epsilon$=0.1) in \pp\ collisions at $\sqrt{s}$~=~7, 10 and 14~TeV (left) in one unit of rapidity at mid-rapidity and expected isolated yields and $p_T$ reach at $\sqrt{s}$~=~7~TeV, for an integrated luminosity of L~=~15~pb$^{-1}$ (right).}
      \label{dndptgamma}
      \end{center}
\end{figure} 
Figure~\ref{dndptgamma} shows the expected isolated photons ($R$~=0.2, $\epsilon$=0.1) cross-sections at 7~TeV and, for comparison, at 10 and 14~TeV (left) and the corresponding yields in \pp\ at 7~TeV (right).
The expected $p_T$ reach, which will be accessible in the ALICE EMCAL acceptance ($\Delta \eta=1.4,\Delta \phi$=1.1), during the first year of LHC running at $\sqrt{s}$~=~7~TeV, 
considering an integrated luminosity\footnote{We assume fully-efficient high-$p_T$ EMCal cluster triggers with no scale-down reduction.} of L~=~15~pb$^{-1}$,  is found to be around 80~GeV/c.
\begin{figure}[htbp!]
\begin{center}
    \includegraphics[height=6.cm,width=7.5cm, clip=true]{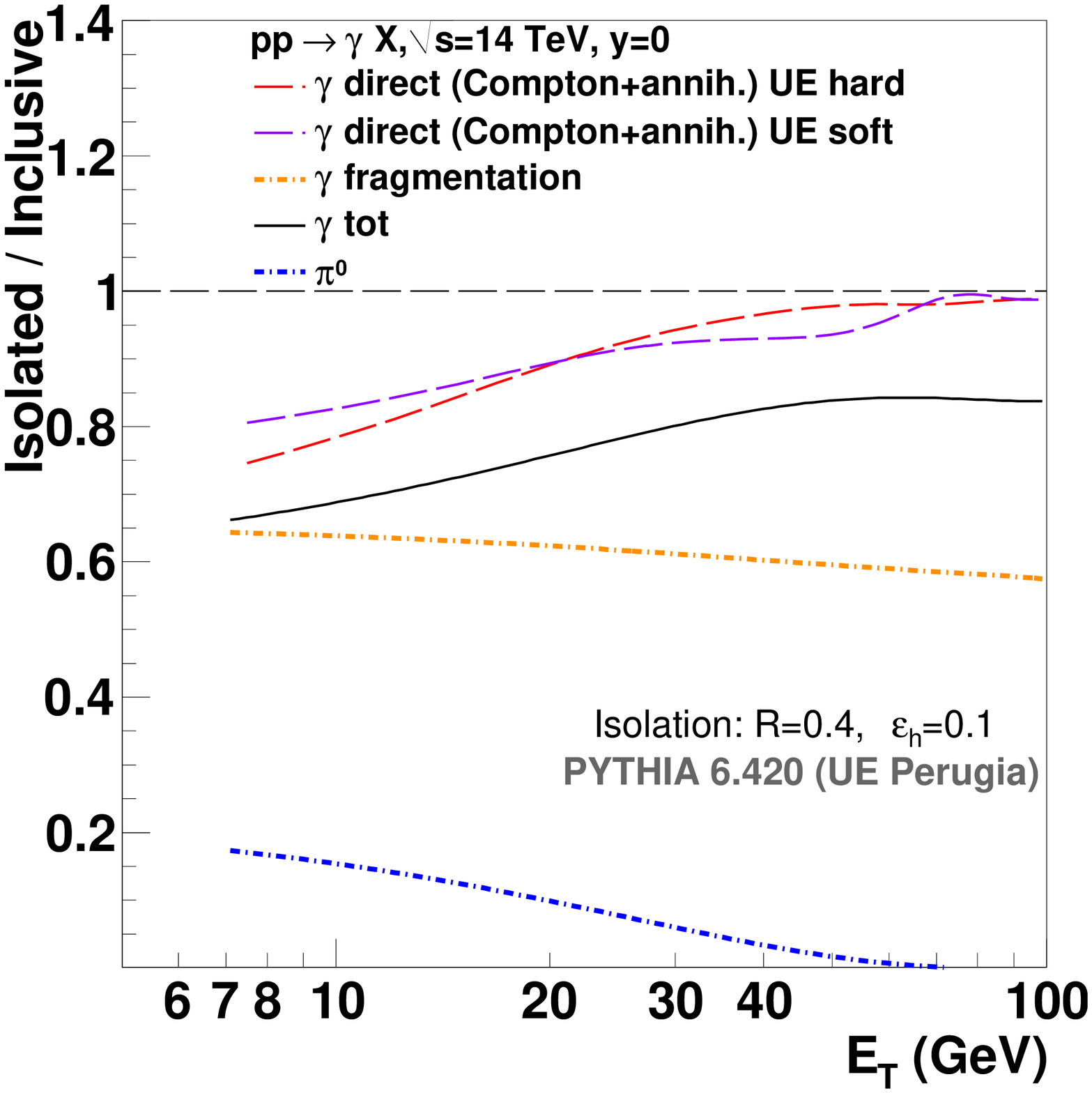}
    \includegraphics[height=6.cm,width=7.5cm, clip=true]{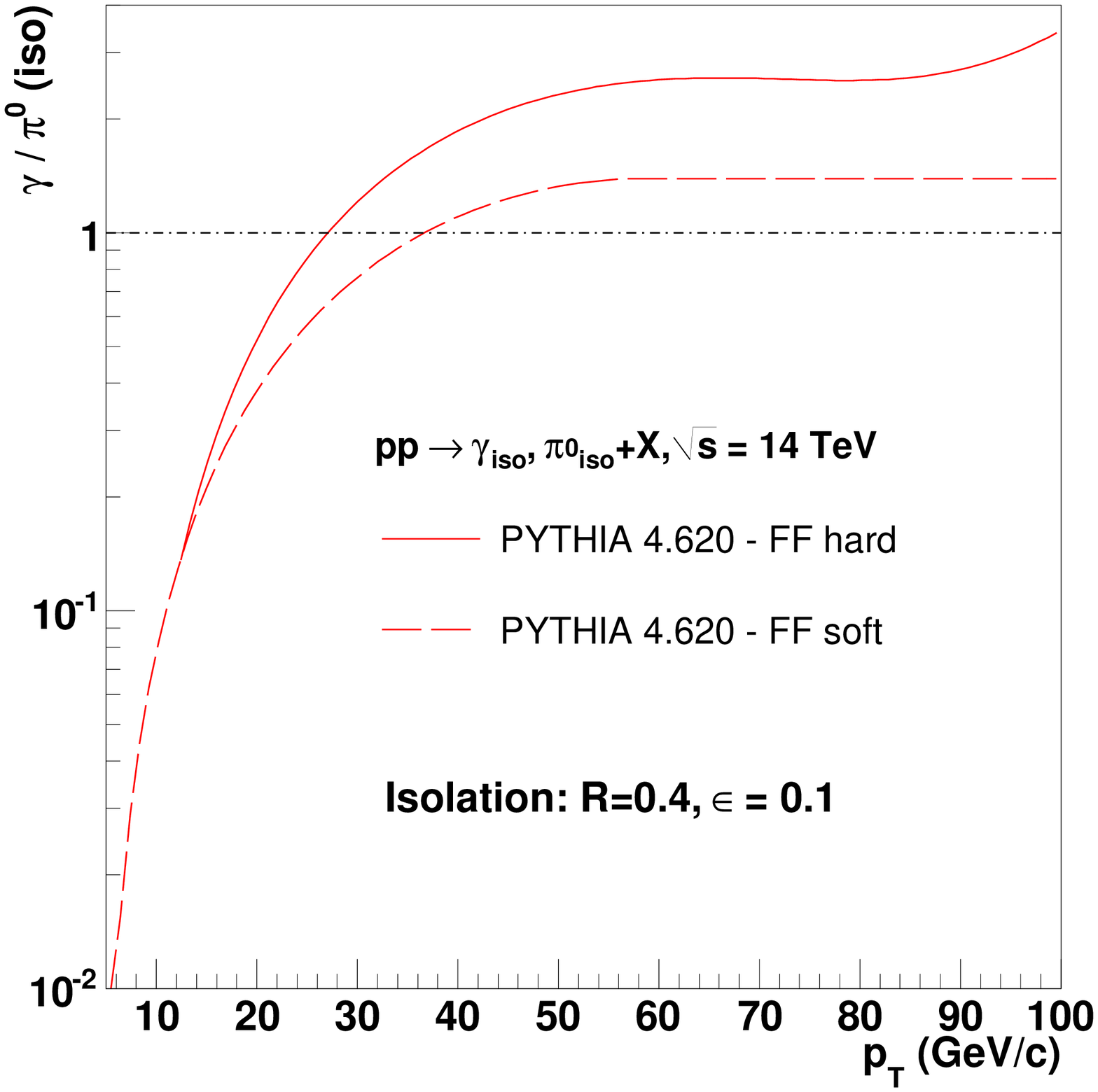}
          \caption{Fraction of isolated ($R$~=~0.4, $\varepsilon$~=~0.1) over inclusive photons and $\pi^0$ for \pp\ collisions simulated by \ttt{PYTHIA} 
          at $\sqrt{s}$ =14 TeV (left), and signal ($\gamma$) over background ($\pi^0$) after isolation cuts for both the FF hard and soft \ttt{PYTHIA} tunes (right).}
      \label{MC}
      \end{center}
\end{figure} 
\ttt{PYTHIA~6.420}~\cite{Sjostrand:2006za}, with the Perugia hard and soft tunes~\cite{Skands:2010ak}, has been used to study the signal and the background, for \pp\ collisions at $\sqrt{s}$~=~14~TeV. 
We have generated the different photon subprocesses: Compton (\ttt{MSUB~=~29}), annihilation (\ttt{MSUB~=~14}) and fragmentation (\ttt{MSEL~=~14}), as well as $\pi^0$ (\ttt{MSEL~=~1}) events, over one unit of rapidity at $y = 0$, for different $p_T$ bins: [5-20], [20-50], [50-100], [100-250], [250-500] and [500-1000]~GeV/c, with 10 millions events in each.
Isolation cuts ($R$~=~0.4, $\varepsilon$~=~0.1) have been applied on them.
Figure~\ref{MC} (left) shows the fraction of isolated over inclusive particles, for the different photon subprocesses and for $\pi^0$, at the MC level.
For the direct $\gamma$, which are theoretically all isolated, we have studied the effect of the UE on the isolation by using two different \ttt{PYTHIA} tunes, hard and soft, which differ by their proportions of initial and final-state radiation, multiple interactions, beam remnants and parton-to-hadron fragmentation functions (FF).
The fraction of isolated direct photons ($R$~=~0.4, $\varepsilon$~=~0.1) goes from 80\% to 100\% from 10 to 100~GeV. Thus, the UE leads to a maximum of 20\% ($\pm$5\%) of isolated direct photons loss at 10~GeV.
The fraction of isolated fragmentation photons with \ttt{PYTHIA} is around 60\%. This leads to a total amount of isolated over inclusive $\gamma$ of about 70-80\% over the full $E_T$ range.
The isolated $\pi^0$ fraction is represented by high-$z$ $\pi^0$ which carry a large fraction of the jet energy ($z$~=~$p_{hadron}/p_{parton}$) and are thus isolated from accompanying hadronic activity. With \ttt{PYTHIA} we find a fraction of isolated $\pi^0$ going from 20\% to 3\% from 10~GeV to 60~GeV. High-$z$ isolated $\pi^0$ are thus considered as the main background in our prompt isolated photon analysis.\\
Figure~\ref{MC} (right) shows the MC-level predictions for the signal over background (S/B) with isolation cuts, for both \ttt{PYTHIA} tunes, in order to assess the systematic uncertainty linked to the choice of the FF. Isolation allows one to strongly enhance the signal over background ratio. Yet, with isolation cuts, the S/B remains smaller than 1 up to $\sim$~30~GeV/c.
In order to suppress the remaining background due to isolated $\pi^0$, one thus needs to employ shower-shape cuts which improve the photon over $\pi^0$ ratio, and subtract statistically any final remaining
$\pi^0$ background.

\section{Analysis}
The goal of this analysis is to obtain a fully corrected isolated prompt photon spectrum at 14~TeV.
The isolated photon cross-section can be written as :
\begin{equation}
E\frac{d^3\sigma}{dp^3} = \frac{1}{2 \pi p_T} \frac{1}{p(\gamma | \gamma)} \left( \frac{ \mbox{N(clusters iso. et id. $\gamma$)}}  {A \times L \times \epsilon \times \Delta p_T \times \Delta \eta} -  p(\gamma | \pi^0)\frac{d\sigma^{\pi^0_{iso} }}{dp_T}    \right)
\label{formule}
\end{equation}
with N(clusters id. $\gamma$ and iso.) the number of clusters identified as photons and isolated, which represents the Signal+Background, p($\gamma | \gamma$) the gamma identification efficiency~\footnote{Defined as a conditionnal probability of identifying a reconstructed cluster as a $\gamma$ knowing that it is indeed a $\gamma$.} and p($\gamma | \pi^0$) the identification contamination due to $\pi^0$ identified as photon, $L$ the integrated luminosity, $A$ the acceptance term, and $\varepsilon$ all the other corrections (reconstruction and UE isolation, shown in Fig.~\ref{MC} left).
For the acceptance term, we consider a fiducial cut corresponding to an isolation radius of $R$~=~0.4, which represents 1/4 of EMCal acceptance. The acceptance with respect to full coverage for one unit of rapidity at mid-rapidity is thus found to be around 10\%.
Full simulation and reconstruction events for signal ($\gamma$-jet) and background (di-jets) at 14~TeV have been used to obtain all correction factors in Eq.~(\ref{formule}).
\begin{figure}[htbp!]
\begin{center}
        \includegraphics[scale=0.38, clip=true]{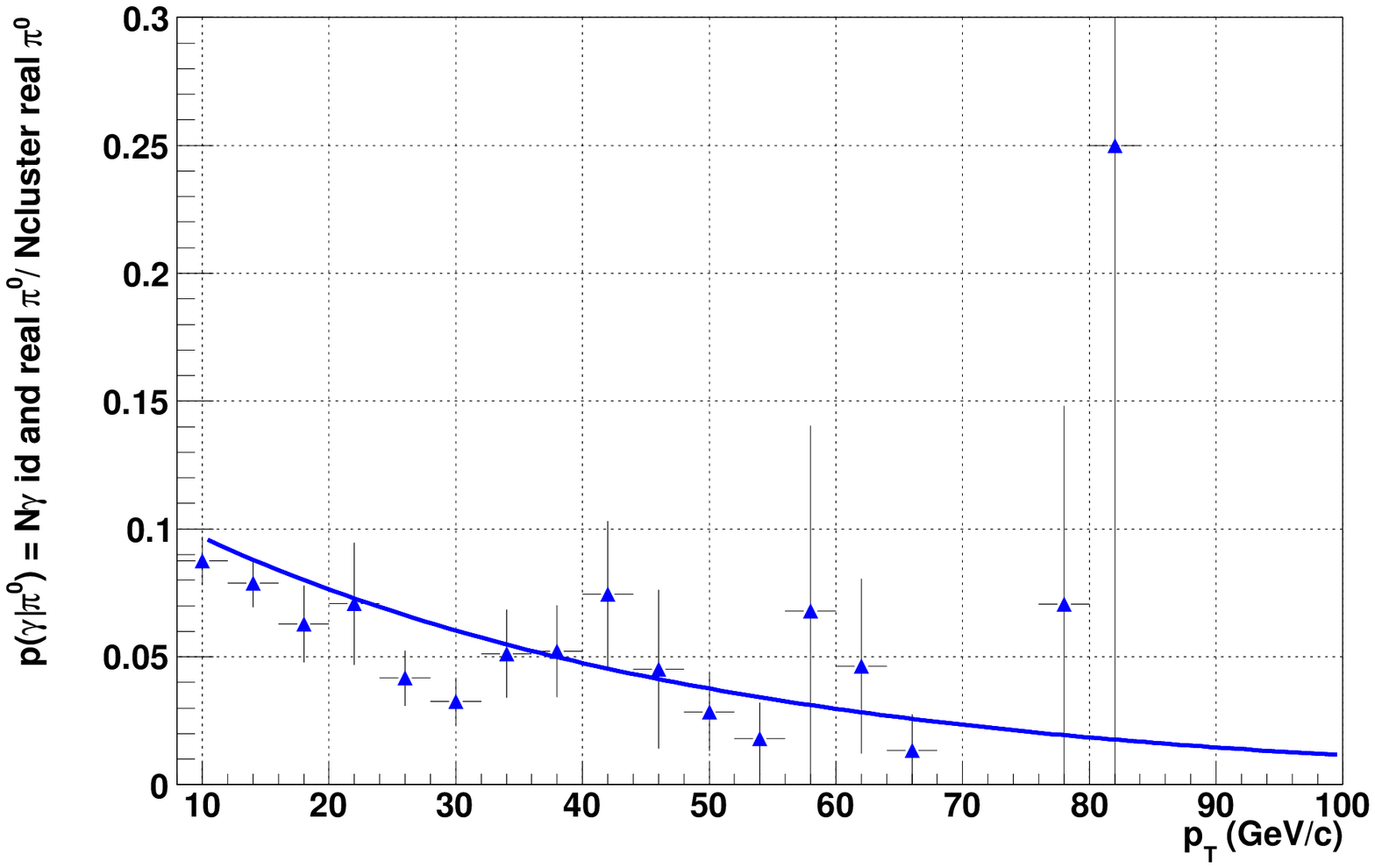}
        \includegraphics[scale=0.34, clip=true]{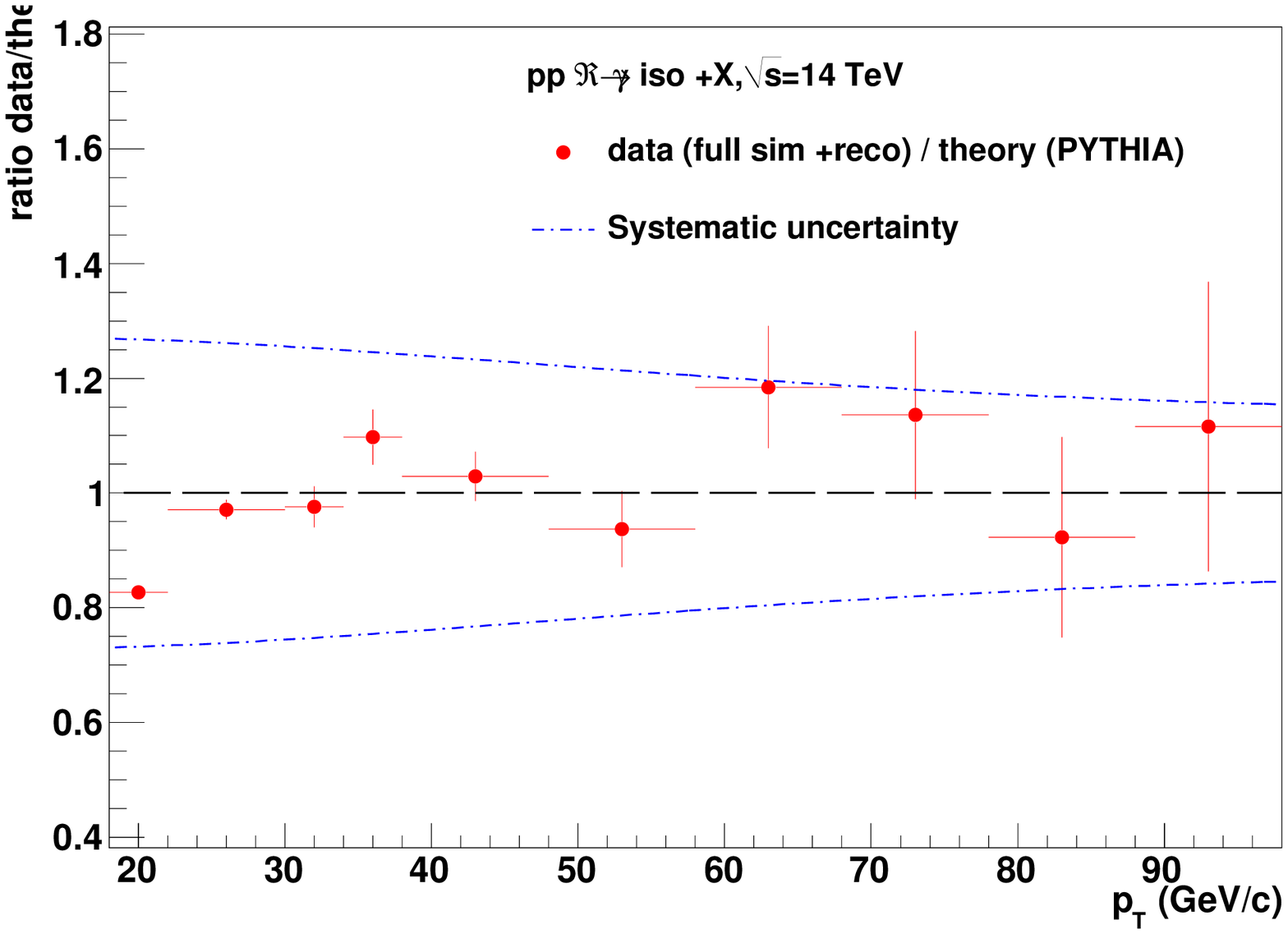}
          \caption{PID contamination of the $\pi^0$ identified as $\gamma$, p($\gamma |\pi^0$) (left), and ratio data/theory (full sim reco vs PYTHIA) for isolated $\gamma$ ($R$~=~0.4, $\varepsilon$~=~0.1) in \pp\ at $\sqrt{s}$ =14 TeV (right).}
      \label{pid}
      \end{center}
\end{figure} 
The clusters are identified by the shower shape Bayesian method~\cite{ppremcal}. A track-matching cluster--charged particle cut, using information from the central tracker, is also applied on the photon candidates.
The photon identification efficiency, $p(\gamma|\gamma)$, is found to
be constant at $\sim$80\% up to 45 GeV/c. The background contamination
due to $\pi^0$'s identified as $\gamma$, $p(\gamma|\pi^0)$, is found to
be less than 10\% above 10 GeV/c and decreasing with $p_T$ (Fig.~\ref{pid} left). Our
shower-shape cuts remove, thus, a significant fraction of neutral pions.
The reconstruction efficiency is $\sim$90\% up to $\sim$80~GeV/c.
The signal/background after photon reconstruction, identification and isolation is found to be greater than one above $p_T\approx$~15 GeV/c.

Finally, we use the \ttt{PYTHIA} spectrum of isolated $\pi^0$ to statistically remove any remaining contamination. 
This $\pi^0$ cross-section, weighted by the term $p(\gamma | \pi^0)$, is subtracted from the fully-corrected left term of Eq.~(1) to
obtain the final corrected isolated $\gamma$ spectrum at 14~TeV. The systematic errors propagated on the final yields are dominated by the EMCal energy
scale (about 15\%) and the (theoretical) uncertainty in the subtracted isolated-$\pi^0$'s ($\sim$20\%). Fig.~\ref{pid} (right) shows the final
data/theory ratio, i.e. the ratio of the full sim+reco spectrum over the original \ttt{PYTHIA} distribution. The fully corrected spectrum is in
good agreement with the input MC LO spectrum within the expected statistical (for 24 pb$^{-1}$) and systematic uncertainties.



\section*{References}

\medskip
\begin{thebibliography}{9}

\bibitem{Aurenche:2006vj}
  P.~Aurenche, M.~Fontannaz, J.~P.~Guillet, E.~Pilon and M.~Werlen,
  Phys.\ Rev.\  D {\bf 73} (2006) 094007
  
\bibitem{Ichou:2010wc}
  R.~Ichou and D.~d'Enterria,
  Phys.\ Rev.\  D {\bf 82} (2010) 014015
  
 \bibitem{Klein:2008di}
  M.~Klein and R.~Yoshida,
  Prog.\ Part.\ Nucl.\ Phys.\  {\bf 61} (2008) 343

 


\bibitem{jetphox}
JetPHOX, P.~Aurenche et al.,
    (\verb$http://lappweb.in2p3.fr/lapth/PHOX_FAMILY/jetphox.html$)

\bibitem{Sjostrand:2006za}
  T.~Sjostrand, S.~Mrenna and P.~Z.~Skands,
  JHEP {\bf 0605}, 026 (2006)

\bibitem{Skands:2010ak}
  P.~Z.~Skands,
  arXiv:1005.3457 [hep-ph].

\bibitem{ppremcal}
    Bellwied, R. et al.,
    arXiv:1008.0413.


\end{thebibliography}
\end{document}